\documentclass{jltp}
\usepackage{amsmath}

\title{Robust Superfluid Phases of $^3$He in Aerogel}

\author{I.A.Fomin\address{P. L. Kapitza
Institute for Physical Problems, \\ul. Kosygina 2, 119334 Moscow,
Russia}}

\runninghead{\bf I.~A.~Fomin}{Robust Superfluid Phases..,}

\begin{document}

\maketitle

\begin{abstract}
  Within a phenomenological approach possible forms of the order parameter
of the superfluid phases of $^3$He in a vicinity of the transition temperature
are discussed. Effect of aerogel is described by a random tensor field
interacting with the orbital part of the order parameter. With respect to
their interaction with the random tensor field a group of "robust" order
parameters which can maintain long-range order in a presence of the
random field is specified. Robust order parameters, corresponding
to Equal Spin Pairing (ESP) states are found and proposed as candidates for
the observed A-like superfluid phase of liquid $^3$He in aerogel.

\end{abstract}


\section {Introduction}
It has been shown before \cite{fom}  that aerogel influences a choice of
superfluid phases of $^3$He. Only  phases which satisfy the condition
  $$
   \eta_{j l}A_{\mu j}A_{\mu l}^*=0.                              \eqno(1)
  $$
can maintain a long-range order. Here $\eta_{j l}=\eta_{j l}({\bf r})$
is an arbitrary real traceless symmetrical tensor field. It describes in
Ginzburg and Landau region a local splitting of the phase transition
temperature for different spherical harmonics with $l=1$.
Condition (1) serves as criterium for a choice of correct zeroth order
combinations of $A_{\mu j}$ on the perturbation $\eta_{j l}({\bf r})$, it
can be rewritten in a form

$$
A_{\mu l}A_{\mu j}^* + A_{\mu j}A_{\mu l}^*=\delta_{jl}\cdot const. \eqno(2)
$$
Let us refer as "robust" the phases for which either condition (1) or
equivalently (2) is met. It can be checked easily that the BW-phase of
pure (free of aerogel) $^3$He is robust. On the other hand ABM-phase
with the order parameter
$$
A_{\mu j}=\Delta\frac{1}{\sqrt{2}}\hat d_{\mu}(\hat m_j+i\hat n_j), \eqno(3)
$$

when substituted in the left-hand side of eq.(2) renders
$\sim -\eta_{jn}l_jl_n\ne 0$ and is not robust (here
${\bf\hat m}$,  ${\bf\hat n}$ and
${\bf\hat l}={\bf\hat m}\times{\bf\hat n}$ are mutually orthogonal unit
vectors). According to general argument of Imry and Ma \cite{imry} in that
case orientation of ${\bf\hat l}$ can be preserved only over a limited
distance, i.e. Cooper pairing can be realized only as a localized state.
In the experiments \cite{osher,dmit} two superfluid phases with different
static and dynamic magnetic properties are observed. One of the phases is
analogous to the BW-phase of pure $^3$He, the other must be of ESP-type, i.e.
for a certain choice of  quantisation axis it
does not contain  component with zero projection of spin. The order parameter
of an ESP-phase can be written as
$$
A_{\mu j}=\Delta\frac{1}{\sqrt{3}}[\hat d_{\mu}( m_j+i n_j)+
                      \hat e_{\mu}( l_j+i p_j)]            , \eqno(4)
$$
For this phase to be robust four vectors {\bf m,n,l,p} must satisfy
the equation
$$
 m_jm_l+n_jn_l+l_jl_l+p_jp_l=\delta_{jl}\cdot const.            \eqno(5)
$$
One  solution of that equation ({\bf p}=0. {\bf m,n,l} - orthonormal
set) was discussed in ref. \cite{fom}. In what follows all solutions
of eq. (5) will be found and a choice of physically relevant solution is
discussed.

\section {Robust ESP phases}
To find all solutions of eq. (5) let us consider 4-dimensional equation
$$
 m_rm_s+n_rn_s+l_rl_s+p_rp_s=\delta_{rs},                        \eqno(6)
$$
where r,s=1,2,3,4. General (up to an overall rotation)
solution of this equation is a set of four orts
$\hat q^{(a)}$ : $\hat q^{(a)}\cdot\hat q^{(b)}=\delta^{ab}$. Now let us fix an
arbitrary four-dimensional unit vector $\hat\nu=(\nu_1,\nu_2,\nu_3,\nu_4)$ and
project  $\hat q^{(a)}$ on a three-dimensional hyperplane, orthogonal
to $\hat\nu$. Four resulting vectors, belonging to this hyperplane
$$
{\bf m}=\hat q^{(1)}-\nu_1\hat\nu, {\bf n}=\hat q^{(2)}-\nu_2\hat\nu,
{\bf l}=\hat q^{(3)}-\nu_3\hat\nu, {\bf p}=\hat q^{(4)}-\nu_4\hat\nu, \eqno(7)
$$
satisfy eq.(6). One can check it  multiplying combination
$m_jm_l+n_jn_l+l_jl_l+p_jp_l$, with {\bf m,n,l,p} expressed according to
eqns. (7),  by an arbitrary vector {\bf a}, orthogonal to $\hat\nu$.
From definitions (7) there follow properties of vectors {\bf m,n,l,p}:
\begin{gather*}
  m^2+n^2+l^2+p^2=3 \\
  {\bf m}\cdot{\bf n}=-\nu_1\nu_2,\quad {\bf m}\cdot{\bf l}=-\nu_1\nu_3,\quad
  {\bf n}\cdot{\bf l}=-\nu_2\nu_3, \ldots \\
  m^2=1-\nu_1^2,\quad n^2=1-\nu_2^2, \ldots
\end{gather*}
With the aid of the second property one can show easily, that
$[{\bf m}\times{\bf n}]\cdot [{\bf l}\times{\bf p}]=0 $.
It means that the normals to the planes spanned correspondingly by the
pairs of vectors {\bf m,n} and {\bf l,p} are mutually orthogonal.
That statement applies to any other choice of pairs among the four
vectors {\bf m,n,l,p}.

So, we conclude that the order parameter  of an arbitrary robust
ESP-phase has a form given by eq. (4) with the vectors {\bf m,n,l,p},
specified by eqns. (7). This is a three-parametric family. Since the
conditions  (1) and (2) have a meaning of  localization threshold,
these order parameters need not be minima of free energy. There may be
localized states with the lower energy i.e. with a higher transition
temperature. Comparison of energies is of importance for a choice within
the three-parameter manifold of the found robust phases.
Zeroth order Ginzburg and Landau free energy
for robust phases has a form:
$$
F_{GL}=\int d^3r\{\alpha(T-T_c)A_{\mu j}A_{\mu j}^*+
      \frac{1}{2}\sum_{n=1}^5 \beta_nI_n\},                         \eqno(8)
$$
where $I_n$ - are invariants of the 4-th order
 and $\beta_1,... \beta_5$ --
phenomenological coefficients (for definitions cf.\cite{vollh}).
Minimization of the free energy (8) is equivalent to minimization of a sum
$\Sigma=\sum_{n=1}^5 \beta_nI_n$. Direct substitution of the found order
parameter renders up to a positive coefficient
$$
\Sigma=\beta_1+9\beta_2+\beta_3+5\beta_4+5\beta_5-
4(\beta_1+\beta_5)(\nu_1\nu_4-\nu_2\nu_3)^2.                      \eqno(9)
$$
Without strong coupling corrections both $\beta_1$ and $\beta_5$ are negative.
It is not likely that the sum $\beta_4+\beta_5$ will change its
sign when these corrections as well as possible corrections due to
a presence aerogel are taken into account. In that case minimum of $\Sigma$ is
achieved when
$$
\nu_1\nu_4=\nu_2\nu_3.                                             \eqno(10)
 $$
This condition has simple physical meaning.  Order parameters
(4) correspond to nonunitary phases, which can have net spin.  Spin density is
proportional to $e_{\mu\nu\lambda}A_{\mu j}A_{\nu j}^*$. For the order
parameter (4) this combination is proportional to
$[\hat d\times\hat e][{\bf n}\cdot{\bf l}-{\bf m}\cdot{\bf p}]$.
With the aid of the listed above properties of {\bf m,n,l,p}  one
can see that the constraint (10) is equivalent to requirement of absence of a
net spin density (or absence of ferromagnetism). Three-dimensional vectors
{\bf m,n,l,p} are determined when the projections $\nu_1,\nu_2,\nu_3,\nu_4$ are
specified.
With the constraint (10) the family of robust ESP-phases has two-parametric
degeneracy.  Possible parametrization is:  $\nu_1=\sin\alpha\sin\beta,
\nu_2=\sin\alpha\cos\beta, \nu_3=\cos\alpha\sin\beta,
\nu_4=\cos\alpha\cos\beta$.
Further lifting of degeneracy can occure in the next order on $\eta_{j l}$,
which is not considered here.

  The ratio of the energy gain of family of robust
phases to that of ABM phase can be expressed in terms of the coefficients
$\beta_1,... \beta_5$:
$$
\frac{\Delta F_{ABM}}{\Delta F_R}=\frac{\beta_{13}+9\beta_2+5\beta_{45}}
{9\beta_{245}},                                                     \eqno(11)
$$
where $\beta_{13}=\beta_1+\beta_3$ etc.. For the weak coupling values of
$\beta_1,... \beta_5$ \cite{vollh} one has
$(\Delta F_{ABM})/(\Delta F_{\nu})$=19/18, i.e. ABM-phase is slightly more
favorable energetically. One has to remark that for the weak coupling
$\beta_1,... \beta_5$ ABM-phase itself is not stable. Combinations of
$\beta_1,... \beta_5$, deduced from experiments \cite{gould} do not form
a complete set and do not define the ratio eq.(11) without ambiguity.

Lifting of the relative spin-orbit  degeneracy
is due to the dipole energy. For the order parameter (4) it has a form:
$$
U_D=\frac{g_D\Delta^2}{3}[({\bf d}\cdot{\bf m}+{\bf e}\cdot{\bf l})^2+
({\bf d}\cdot{\bf n}+{\bf e}\cdot{\bf p})^2-
({\bf d}\cdot{\bf l}-{\bf e}\cdot{\bf m})^2-
({\bf d}\cdot{\bf p}-{\bf e}\cdot{\bf n})^2].               \eqno(12)
$$
This energy reaches its minimum at ${\bf d}\cdot{\bf m}=0,
{\bf e}\cdot{\bf l}=0, {\bf d}\cdot{\bf n}=0, {\bf e}\cdot{\bf p}=0,
{\bf e}\cdot{\bf m}=-{\bf d}\cdot{\bf l},
{\bf e}\cdot{\bf n}=-{\bf d}\cdot{\bf p}.$
These conditions fix orientation of the orbital vectors {\bf m,n,l,p} with
respect to the spin vectors {\bf d,e} and ${\bf f}={\bf d}\times{\bf e}$.
In a coordinate system determined by a reper ({\bf d,e,f}) the orbital vectors
have coordinates:
${\bf m}=(0,-\cos\beta,\cos\alpha\sin\beta);
{\bf n}=(0,\sin\beta,\cos\alpha\cos\beta);
{\bf l}=(\cos\beta,0,-\sin\alpha\sin\beta);
{\bf p}=(-\sin\beta,0,-\sin\alpha\cos\beta)$,
where $\alpha$ and $\beta$ are the same as in the above parametrisation of
$\nu_1,\nu_2,\nu_3,\nu_4$.

\section {Discussion}

Eqation (4) with the constraint (5) determines a two-parametric
family of order parameters for a triplet {\it p}-wave Cooper pairing. These
order parameters can preserve long-range order in a presence of the random
tensor field modelling aerogel in the superfluid $^3$He. Magnetic
susceptibility of the phases corresponding to that type of the order parameter
is that of the normal phase, so they can be  considered as candidates for the
A-like phase of $^3$He in aerogel. Calculation of other observable properties
of these phases for definitive identification of the A-like phase is in a
progress.

 This research was supported in part by RFBR grant (no. 01-02-16714) and by
 NATO grant SA(PST.CLG.979379)6993/FP.


\begin{thebibliography}{99}

\bibitem{fom} I.A. Fomin, {\it JETP Letters},   {\bf 77}, 240 (2003)

\bibitem{imry} Y. Imry and S. Ma {\it Phys. Rev. Lett.},
               {\bf 35}, 1399 (1975)
\bibitem{osher} B.I.Barker, Y.Lee, L.Polukhina
{\it et al. Phys. Rev. Lett.}, {\bf 85}, 2148 (2000)

\bibitem{dmit} V.V. Dmitriev, V.V.Zavjalov, D.E.Zmeev, I.V. Kosarev,
               and N. Mulders, {\it JETP Letters},  {\bf 76}, 321 (2002).

\bibitem{vollh} D. Vollhardt and P. W\"olfle, {\it The Superfluid Phases
of Helium 3}, Taylor and Francis  (1990).

\bibitem{gould} H.M. Bozler and C.M.Gould, {\it Czech. J. Phys., Suppl S1},
               {\bf 46}, 165 (1996)


 \end{thebibliography}
  \end{document}